\documentclass[
    ,final            
  ]
  {aipproc}

\usepackage{amsmath,amssymb}

\layoutstyle{6x9}

\begin{document}

\title{GRB 050814 at $z = 5.3$ and the Redshift Distribution of 
\emph{Swift} GRBs}


\author{P.~Jakobsson}{
  address={Dark Cosmology Centre, Niels Bohr Institute, University of
          Copenhagen, Juliane Maries Vej 30, 2100 Copenhagen, Denmark}
}

\author{A.~Levan}{
  address={Centre for Astrophysics Research, University of Hertfordshire,
          College Lane, Hatfield, Herts, AL10 9AB, UK}
}

\author{J.~P.~U.~Fynbo}{
  address={Dark Cosmology Centre, Niels Bohr Institute, University of
          Copenhagen, Juliane Maries Vej 30, 2100 Copenhagen, Denmark}
}

\author{R.~Priddey}{
  address={Centre for Astrophysics Research, University of Hertfordshire,
          College Lane, Hatfield, Herts, AL10 9AB, UK}
}

\author{J.~Hjorth}{
  address={Dark Cosmology Centre, Niels Bohr Institute, University of
          Copenhagen, Juliane Maries Vej 30, 2100 Copenhagen, Denmark}
}

\author{N.~Tanvir}{
  address={Centre for Astrophysics Research, University of Hertfordshire,
          College Lane, Hatfield, Herts, AL10 9AB, UK}
}

\author{D.~Watson}{
  address={Dark Cosmology Centre, Niels Bohr Institute, University of
          Copenhagen, Juliane Maries Vej 30, 2100 Copenhagen, Denmark}
}

\author{B.~L.~Jensen}{
  address={Dark Cosmology Centre, Niels Bohr Institute, University of
          Copenhagen, Juliane Maries Vej 30, 2100 Copenhagen, Denmark}
}

\author{J.~Sollerman}{
  address={Dark Cosmology Centre, Niels Bohr Institute, University of
          Copenhagen, Juliane Maries Vej 30, 2100 Copenhagen, Denmark}
}

\author{P.~Natarajan}{
  address={Department of Astronomy, Yale University, PO Box 208101,
          New Haven CT 06520-8101, USA}
}

\author{J.~Gorosabel}{
  address={Instituto de Astrof\'{\i}sica de Andaluc\'{\i}a (CSIC),
          Apartado de Correos, 3004, E-18080 Granada, Spain}
}

\author{J.~M.~Castro~Cer\'on}{
  address={Dark Cosmology Centre, Niels Bohr Institute, University of
          Copenhagen, Juliane Maries Vej 30, 2100 Copenhagen, Denmark}
}

\author{K.~Pedersen}{
  address={Dark Cosmology Centre, Niels Bohr Institute, University of
          Copenhagen, Juliane Maries Vej 30, 2100 Copenhagen, Denmark}
}

\begin{abstract}
We report optical, near-infrared and X-ray observations of the afterglow of 
GRB\,050814, which was seen to exhibit very red optical colours. By modelling 
its spectral energy distribution we find that $z = 5.3 \pm 0.3$. We next
present a carefully selected sample of 19 \emph{Swift} GRBs, intended to 
estimate in an unbiased way the GRB redshift distribution, including the mean 
redshift ($z_\mathrm{mean}$) as well as constraints on the fraction of 
high-redshift bursts. We find that $z_\mathrm{mean} = 2.7$ and that at least 
5\% of the GRBs originate at $z > 5$. The redshift distribution of the sample 
is qualitatively consistent with models where the GRB rate is proportional 
to the star formation rate in the Universe. The high mean redshift of this 
GRB sample and the wide redshift range clearly demonstrates the suitability 
of GRBs as efficient probes of galaxies and the intergalactic medium over a 
significant fraction of the history of the Universe.
\end{abstract}

\maketitle


\section{Introduction}
The immense luminosities of the gamma-ray bursts (GRBs), coupled with their 
origin in the core collapse of massive stars \cite{jens,stanek} and their
$\gamma$-ray penetration through dust, open up a variety of intriguing 
cosmological applications. Much effort has been directed into the 
use of GRBs for studying star formation (e.g. \cite{lise}), as backlights for 
exploring high-redshift galaxies and the intergalactic medium 
(e.g. \cite{paul,palli04}), and even as probes of cosmological parameters 
(e.g. \cite{ghirlanda,mortsell}). Although the GRB population observed until 
the end of 2004 had enabled much progress in the field, it was widely expected 
that the launch of {\it Swift}, and the subsequent order of magnitude 
increase in the number of GRBs open to detailed study, would allow further 
insight into the high-redshift Universe \cite{gehrels}. Indeed, the ability 
of {\it Swift} to locate and follow-up a fainter burst population than was 
previously possible \cite{edo05} has allowed the study of more distant bursts. 
The mean redshift of pre-{\it Swift} bursts was $z_\mathrm{mean} = 1.4$, 
while we show here that bursts discovered by {\it Swift} now have 
$z_\mathrm{mean} = 2.7$, including the first burst to have been discovered 
with $z > 6$, GRB\,050904 at $z=6.295$ (e.g. \cite{kawai}).
\section{Spectral energy distribution of the afterglow}
\label{sed.sec}
\begin{figure}
\centering
\includegraphics[width=0.8\textwidth]{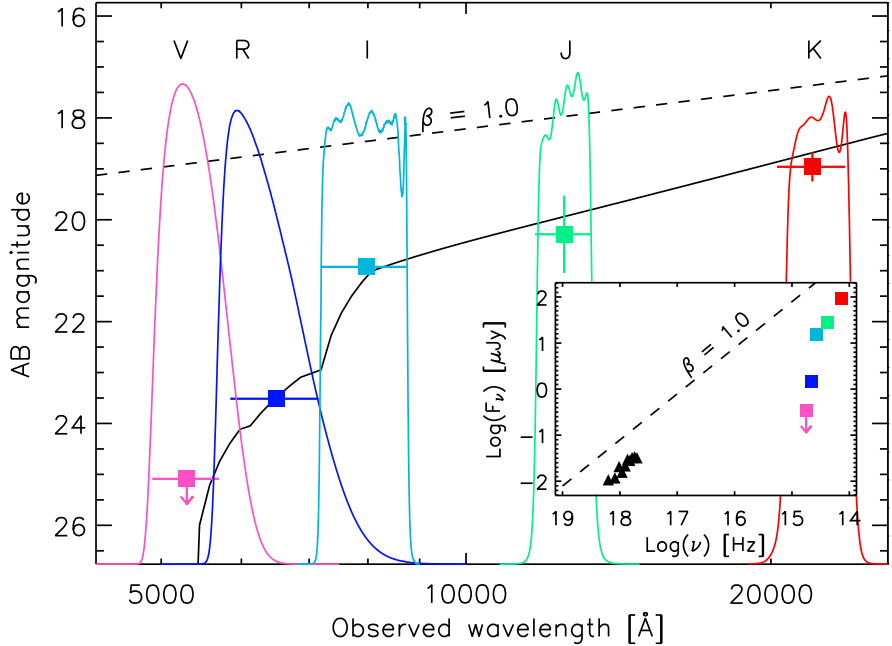}
\caption{The spectral energy distribution of the GRB 050814 afterglow at
$\Delta t = 14$\,hr. The strong break blueward of the $I$-band is 
too strong to be readily explained by reddening alone and is best fit by 
the presence of the Ly$\alpha$ break at $z = 5.3$. The solid curve is a
fit to the data at that redshift. The dashed line shows the spectral 
slope expected from a synchrotron emission in the fireball
model with $\beta = 1$. The filter response functions are also shown.
The horizontal error bars represent the FWHM of each filter. The
$V$-band upper limit is 2$\sigma$. The inset shows the $VRIJK$ observations
(filled squares) along with the X-ray spectrum (filled triangles) at
$\Delta t = 14$\,hr. The dashed line is the same $\beta = 1$ slope as
in the main panel.}
\label{sed.fig}
\end{figure}

Our multiband observations of the GRB\,050814 afterglow are presented in
\cite{palliz}. They allowed the construction of its spectral energy 
distribution (SED), displayed in Fig.~\ref{sed.fig}, where we have corrected 
the observed data points for foreground (Galactic) extinction. The SED has a 
strong break blueward of the $I$-band, exhibiting colours of 
\mbox{$I - K = 3.44 \pm 0.29$\,mag} and \mbox{$R - I = 2.87 \pm 0.10$\,mag}, 
corresponding to spectral slopes of \mbox{$\beta_{IK} = 1.78 \pm 0.12$} and 
\mbox{$\beta_{RI} = 11.70 \pm 0.04$}, respectively ($F \propto \nu^{-\beta}$). 
The latter value is unreasonable for GRB afterglows, implying an electron 
energy power-law index more than ten times higher than normally observed. 
Even in the case of high local extinction ($A_V$), such steep slopes cannot 
be obtained (see also \cite{reichart}). 
\par
The most likely explanation for the steep break observed is due to the 
presence of the Ly$\alpha$ break at a redshift of $5 < z < 6$. To provide a 
more robust estimate of the GRB\,050814 redshift we fit the available 
photometry at different redshifts, allowing for a range in $\beta$ and
$A_V$ modelled using the parametrization of \cite{calzetti}. The models of 
\cite{madau} provide the average hydrogen opacity as a function of redshift. 
The minimum $\chi^2$ is obtained for $z = 5.3 \pm 0.3$. However, we are only 
able to obtain weak constraints on $\beta$ and $A_V$. Fixing $\beta = 1.0$, 
a typical value for GRB afterglows, results in a best fit of a restframe 
$A_V = 0.9$\,mag and an unchanged redshift. This $A_V$ is marginally higher 
than has been inferred from the SEDs of pre-\emph{Swift} bursts \cite{kann} 
with bright optical afterglows (OAs), but is a necessary consequence of the 
red $I - K$ colour. The extrapolated $\beta = 1.0$ line, normalized for 
$A_V = 0.9$\,mag, slightly overestimates the predicted X-ray flux (inset 
of Fig.~\ref{sed.fig}), indicating that $\beta$ is a bit steeper; 
$\beta = 1.1$ would make the X-ray data fall on the line. Since the best 
fit X-ray spectral index is consistent with the assumed optical/NIR one, a 
cooling break between the optical and X-rays can be ruled out.

\section{The Redshift Distribution of \emph{Swift} bursts}
\begin{table}
\centering
\setlength{\arrayrulewidth}{0.8pt}   
\begin{tabular}{@{}lccrrrr@{}}
\hline
\hline
GRB & $z$ & $A_V^\mathrm{Gal}$ & $\theta_\mathrm{Sun}$ & 
$\theta_\mathrm{Moon}$ & $I_\mathrm{Moon}$ & Ref. \\
    &     & [mag] & [deg]                  & [deg]               & [\%] & \\
\hline
051001  &        & 0.05 & 142 & 156 &  4 & \\
051006  &        & 0.22 &  83 & 121 & 12 & \\
051016A &        & 0.29 &  76 & 116 & 98 & \\
051016B & 0.94   & 0.11 &  73 & 117 & 99 & \cite{soderberg} \\
051117B &        & 0.18 & 130 &  48 & 97 & \\
060108  & $<8.5$ \hspace{3.9mm} & 0.05 & 146 &  96 & 69 & \cite{monfardini} \\
060111A & $<5.0$ \hspace{3.9mm} & 0.09 &  61 & 111 & 90 & \cite{blustin} \\
060115  & 3.53   & 0.44 & 121 &  72 & 99 & \cite{piranomonte} \\
060124  & 2.30   & 0.44 & 121 & 132 & 30 & \cite{mirabal} \\
\hline
\end{tabular}
\caption{An update to the list of 28 long-duration GRBs from table~2 in
\cite{palliz}; these additional nine bursts were detected after 30 September 
2005. Here $\theta_\mathrm{Sun}$ is the Sun-to-field distance, 
$\theta_\mathrm{Moon}$ the Moon-to-field distance and $I_\mathrm{Moon}$ the 
Moon illumination at the time the burst occurred. For a burst detected in 
the optical but without a reported redshift, an upper redshift limit is 
estimated based on the filter it is detected in.}
\label{sample.tab}
\end{table}     
In order to study the redshift distribution of GRBs, it is important to 
carefully select a sample containing bursts which have ``observing 
conditions'' favorable for redshift determination. In \cite{palliz} we
introduced four criteria with the aim of constructing such a sample. Here we
recap those criteria and add a fifth one: (1) Small error circles, hence 
the bursts have to be localised with the XRT. (2) The Galactic extinction 
in the direction to the burst has to be sufficiently small or 
$A_V^\mathrm{Gal} < 0.5$\,mag. (3) The XRT error circle should be distributed 
quickly (within 12 hours) for a relatively rapid follow-up. Although the 
automatic slewing of {\it Swift} was enabled in the middle of January 2005, 
part of the following month was dedicated to calibration which could not 
be interrupted. Therefore, we have only included bursts occurring after 
1 March 2005. (4) Rejection of bursts with a declination unsuitable 
(above $+70^{\circ}$ or below $-70^{\circ}$) for follow-up observations.
(5) The Sun-to-field distance has to be large enough, with 
$\theta_\mathrm{Sun} \gtrsim 55^{\circ}$.
\par
The first 28 bursts in the sample were listed in table~2 in \cite{palliz}.
Nine additional bursts are presented in Table~\ref{sample.tab}. For each 
burst we have also listed the Moon-to-field distance ($\theta_\mathrm{Moon}$) 
and the Moon illumination at the time of the burst. This is done to examine 
if these parameters affect the redshift determination significantly, e.g. a 
full Moon close to a burst location. This is of course difficult to quantify 
as the OA brightness also plays a role. For example, GRB\,050820A (table~2 
in \cite{palliz}) has a measured redshift even if it occurred during a full 
Moon and $\theta_\mathrm{Moon} = 34^{\circ}$. Therefore, we decided not to 
limit the sample further. This relatively ``clean'' sample of 37 bursts has 
a redshift recovery rate of roughly 50\% (19/37).
\par
\begin{figure}
\centering
\includegraphics[width=0.8\textwidth]{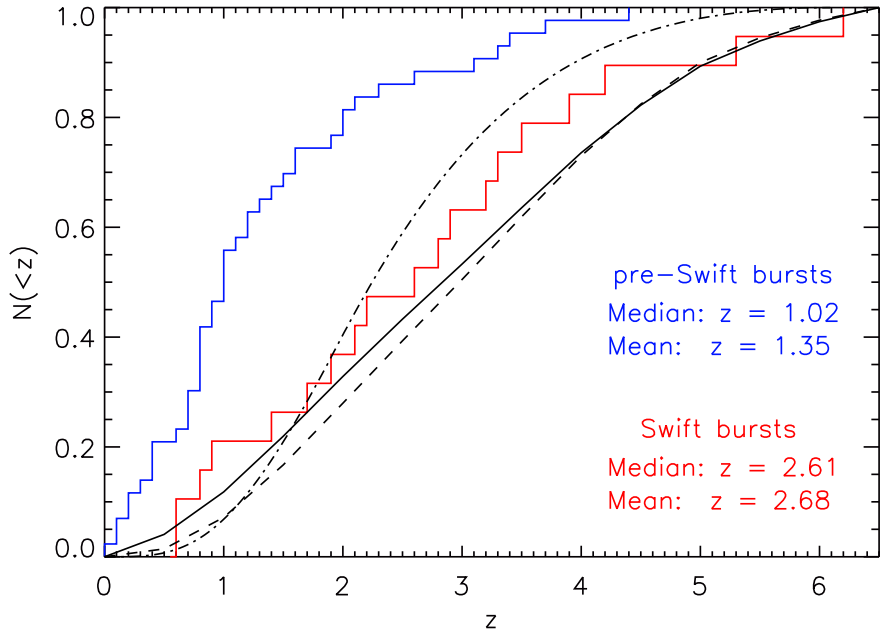}
\caption{The cumulative fraction of GRBs as a function of redshift for 
43 pre-{\it Swift} bursts (upper stepwise curve) and 19 {\it Swift} bursts 
(lower stepwise curve). Overplotted are three simple models for the 
expectation of the redshift distribution of GRBs: model II from \cite{priya} 
in which the GRB rate is proportional to the star formation rate (solid 
curve), model IV from \cite{priya} in which the GRB rate increases with 
decreasing metallicity (dashed curve) and a model from \cite{javier} in 
which the GRB rate is proportional to the star formation rate (dash-dotted 
curve). All three models fold in the \emph{Swift}/BAT flux sensitivity.}
\label{z.fig}
\end{figure}

Figure~\ref{z.fig} shows the redshift distribution of the 19 bursts with a 
reported redshift in our {\it Swift} sample. Both the mean and the median 
is $z \approx 2.65$, more than twice as large as the corresponding numbers 
for pre-{\it Swift} bursts. A natural explanation for this increase is the 
lower trigger threshold of {\it Swift} compared to previous missions, giving 
rise to fainter ({\it Swift} events are on average 1.7\,mag fainter in the 
$R$-band at a similar epoch: \cite{edo05}) and higher redshift bursts. This 
is complemented by the accurate positions provided by {\it Swift} and the 
rapid response of a variety of telescopes aimed at redshift determinations. 
\par
This {\it Swift} sample is the most uniform to date and it is of interest
to compare its redshift distribution to models predicting the fraction of 
GRBs expected to occur at a given redshift. \cite{priya} have modelled the 
expected redshift distribution for GRBs, utilising several models including 
those which follow the globally averaged star formation rate (model II), and 
those which scale according to the average metallicity of the Universe at a 
given redshift (model IV, see e.g. \cite{johan,fryer}). \cite{javier} have 
also carried out a similar exercise, where the GRB rate is assumed to be 
proportional to the star formation rate. These models are plotted in 
Fig.~\ref{z.fig}.
\par
It is remarkable how similar the observed \emph{Swift} redshift distribution 
is to the model predictions; we can now reason that GRBs indeed trace star 
formation (see also \cite{dai,palli05}). However, with the available sample 
and the limited flux sensitivity of the \emph{Swift}/BAT for $z > 5$ bursts, 
it is currently not possible to determine if GRBs are unbiased tracers of 
star formation. For example, models II and IV from \cite{priya} are nearly 
indistinguishable when comparing to the relatively small sample of 19 bursts. 
Note that although model II from \cite{priya} and the \cite{javier} model 
both presuppose that the GRB rate is proportional to the star formation rate, 
they use different assumptions regarding the poorly determined GRB luminosity 
function and the intrinsic spectral shape, explaining their difference in 
Fig.~\ref{z.fig}. 
\par
By including all the bursts in Table~\ref{sample.tab} and table 2 in 
\cite{palliz}, we are able to constrain the number of bursts above a specific 
redshift. For example, 5\%--40\% of the bursts are located at $z > 5$. The 
\cite{priya} and \cite{javier} predictions are 10\% and 2\%, respectively, 
suggesting that the GRB luminosity function parameters and/or the GRB 
spectral index assumed in \cite{priya} might be more appropriate. 
\cite{bromm} also predict that 10\% of the \emph{Swift} GRBs should 
originate at $z > 5$.
\section{Discussion \& Conclusions}
\label{dis.sec}
The mean redshift of our relatively unbiased {\it Swift} sample 
($z_\mathrm{mean} = 2.7$) is larger than the median redshift of sub-mm 
galaxies ($z_\mathrm{median} = 2.2$: \cite{chapman}) and is similar to that 
of Type 2 AGNs ($z_\mathrm{mean} \sim 3$: \cite{padovani}). With two 
$z > 5$ GRBs discovered within a space of a month, and primarily due to the 
spectroscopic redshift of $z = 6.295$ for GRB\,050904 \cite{kawai}, we are 
finally accessing the GRB high-redshift regime. Are we starting to probe 
the era of Pop III stars? If the transition between the dark ages and the 
era of reionization occurred around $z \approx 6$--7 (see e.g. \cite{miralda} 
for a review), the answer might be positive. However, \cite{abel} have 
calculated that at most one massive metal-free star forms per pre-galactic 
halo, and since the GRB progenitor may need to be a member of a close binary 
system in the collapsar scenario (e.g. \cite{fryer99,MacF,zhang}),
it seems unlikely that Pop III stars could end their lives as GRBs. 
\cite{woosley} have proposed a non-binary possibility in the collapsar 
scenario, introducing unusually rapidly rotating massive stars. It is 
therefore possible that Pop III stars are GRB progenitors, although the 
number of unknowns is currently too large to arrive at a concrete conclusion.
\par
The sample contains GRB\,050814, whose OA was particularly faint in the 
$R$-band; the observed optical to X-ray spectral slope is flatter 
($\beta_\mathrm{OX} = 0.36$) than expected for the fireball model. Hence, 
GRB\,050814 is classified as a dark burst as defined by \cite{palliDARK}. 
We have argued that this is most likely due to the 
high-redshift nature ($z = 5.3$) of this burst; the $R - I$ colour is 
extremely red which is impossible to explain by strong extinction given 
the observed $I - K$ colour. Indeed, a similar conclusion was proposed 
for GRB\,980329 \cite{andyZ}.
\par
It is clear that GRBs have now opened up a window to the very high-redshift 
Universe. The emerging GRB redshift histo\-gram (Fig.~\ref{z.fig}) strongly
indicates that GRBs can be used to trace the star formation in the 
Universe over a wide redshift range ($0 \lesssim z \lesssim 7$). Future 
instrumentation, such as the X-shooter \cite{xshooter}, will hopefully shed 
light on the end of the dark ages and the possible GRB/Pop III connection.


\begin{theacknowledgments}
PJ, JF, BLJ and KP acknowledge support from the Instrument Centre for Danish 
Astrophysics (IDA). AL, RP and NT acknowledges PPARC for support. JS 
acknowledges Danmarks Nationalbank for lodging. The research 
of JG is supported by the Spanish Ministry of Science and Education through 
programmes ESP2002-04124-C03-01 and AYA2004-01515. JMCC gratefully 
acknowledges partial support from IDA and the NBI's International Ph.D. 
School of Excellence. The Dark Cosmology Center is funded by the Danish 
National Research Foundation. The authors acknowledge benefits from 
collaboration within the EU FP5 Research Training Network ``Gamma-Ray Bursts: 
An Enigma and a Tool".
\end{theacknowledgments}



\bibliographystyle{aipprocl} 




\end{document}